\documentclass[review,useAMS,usenatbib,referee]{biom}
\newcommand{\Rr}{R_{\boldsymbol \rho}}
\newcommand{\bt}{\boldsymbol \psi}
\usepackage{color, xcolor, colortbl}
\usepackage[figuresright]{rotating}
\usepackage{amsmath, amssymb}
\usepackage{url}
\usepackage{booktabs}

\title[Quadratic Exponential Logistic Regression]{An Accurate Standard Error Estimation for Quadratic Exponential Logistic Regressions by Applying Generalized Estimating Equations to Pseudo-Likelihoods}

\author{Wei Yong Ong \\
	   Department of Biostatistics \& Health Data Science, University of Minnesota School of Public Health, \\ Minneapolis, MN, United States 
	\and 
        Shao-Man Lee \\
          Miin Wu School of Computing, National Cheng Kung University, Tainan City, Taiwan 
        \and
        Chia-Ming Hsueh \\
            Department of International Business and Foreign Languages, \\ Minghsin University of Science and Technology, Hsinchu City, Taiwan
        \and
        Sheng-Mao Chang$^{*}$\email{smchang110@gm.ntpu.edu.tw}\\
        Department of Statistics, National Taipei University, New Taipei City, Taiwan} 
        
\begin{document}

\makeatletter
\renewcommand{\@journal}{}
\renewcommand{\@doi}{}
\renewcommand{\@pagerange}{}
\makeatother

\pagerange{\pageref{firstpage}--\pageref{lastpage}} 


\label{firstpage}


\begin{abstract}
\normalsize

For a set of binary response variables, conditional mean models characterize the expected value of a response variable given the others and are popularly applied in longitudinal and network data analyses. The quadratic exponential binary distribution is a natural choice in this context. However, maximum likelihood estimation of this distribution is computationally demanding due to its intractable normalizing constant, while the pseudo-likelihood, though computationally convenient, tends to severely underestimate the standard errors. In this work, we investigate valid estimation methods for the quadratic exponential binary distribution and its regression counterpart. We show that, when applying the generalized estimating equations to the pseudo-likelihood, using the independence working correlation yields consistent estimates, whereas using dependent structures, such as compound symmetric or autoregressive correlations, may introduce non-ignorable biases. Theoretical properties are derived, supported by simulation studies. For illustration, we apply the proposed approach to the carcinogenic toxicity of chemicals data and the constitutional court opinion wringing data. 

\end{abstract}

%
%

\begin{keywords}
\normalsize
Asymmetric Ising model; Boltzmann machine; Conditional mean model; Markov model; Network data; Transition model.
\end{keywords}

\maketitle

\section{Introduction}\label{sec1}
For a set of correlated binary response variables, a conditional mean model is described as the mean of one variable given all or part of the other variables. Conditional mean models are popularly applied in, but not limited to, longitudinal data and network data. In the context of longitudinal data, the mean model for a current response is influenced by previous responses. Transition models \citep{Agresti2019} with Markov property \citep{Zeger1988} exemplify this under the generalized linear model (GLM; \citealp{McCullagh1983}) framework. For network data \citep{Strauss1990}, especially the non-directed graphs, the log-linear model \citep{Bishop1975} is widely used. Its data analysis majorly relies on modeling one node conditional on the rest.  As a simpler version of the log-linear model, the quadratic exponential binary distribution (QEBD; \citealp{Cox1972, Zhao1990, Cox1994}), also referred to as the asymmetric Ising model \citep{Ravikumar2010} or Boltzmann machine, is gaining interest due to its parallels with the Gaussian distribution. 

The following is a short introduction to the QEBD. Assume that we have $n$ independent random vectors. The $k$th random vector is denoted as $\mathbf Y_k = (Y_{k1},\dots, Y_{km})^{\top}$ where each of $Y_{kj}$'s takes values 0 or 1. Let $\mathbf y_k = (y_{k1},\dots,y_{km})^{\top}$ be a realization of $\mathbf Y_k$. Denote the collection of all possible configurations of $\mathbf Y_k$ as $B^m$. The size of $\mathcal B^m$ is $2^m$ since $Y_{kj}$'s are binary.  The QEBD has the form
\begin{equation} \label{eq:Asy_Ising}
    \Pr(\mathbf Y_k =\mathbf y_k)=\exp\left\{ \sum_{j=1}^m y_{kj}\beta_j + \sum_{1\leq j_1 < j_2\leq m}\theta_{j_1j_2}y_{kj_1}y_{kj_2} - \Lambda \right\}, \quad k=1,\dots, m,
\end{equation}
where $\Lambda = \log \left(\sum_{\mathbf y \in \mathcal B^m}\exp\{\sum_{j=1}^m y_{j}\beta_j + \sum_{j_1 < j_2}\theta_{j_1j_2}y_{j_1}y_{j_2}\}\right)$ is the normalizing constant. Analogous to conventional linear models, $\beta_j$ can be viewed as the ``main effect" of $Y_{kj}$ and $\theta_{j_1j_2}$ represents the ``interaction effect" between $Y_{kj_1}$ and $Y_{kj_2}$.  In terms of the non-directed graph, ``$\theta_{j_1j_2}=0$" means that there is no edge between node $j_1$ and node $j_2$. 

For the estimation aspect, finding the maximum likelihood estimate (MLE) of QEBD can be computationally intensive due to the evaluation of its normalizing constant $\Lambda$ in (\ref{eq:Asy_Ising}). For small $m$'s, according to our simulation studies, the average computing times of the MLE method for $m =$ 5, 10, and 12 are 0.652, 26.346, and 196.759 seconds, respectively. For $m=15$, the computing time of analyzing a single dataset exceeds one hour. The computing time grows exponentially in $m$. For large $m$, classical solutions have been reviewed in \cite{Hastie2009}. Approximating $\Lambda$ plays the central role in estimation. Popular approximations are the iterative proportional fitting \citep{Jirousek1995}, mean field approximation \citep{Peterson1987}, and Gibbs sampling \citep{Ripley1996}. Exploring the MLE via these approximation approaches is computationally expensive, too. Moreover, the biases due to these approximations are unavoidable.

Alternatively, the pseudo-likelihood approach (PL; \citealp{Strauss1990}) mimics the QEBD distribution (\ref{eq:Asy_Ising}) likelihood by the product of conditional distributions. This is achieved by the fact that the conditional distribution of the QEBD can be expressed in the form of logistic regression, 
\[
    \mbox{logit}\left(\Pr(Y_{kj}=y_{kj}| \mathbf Y_{k[j]}= \mathbf y_{k[j]})\right) = \beta_j + \sum_{s \neq j} \theta_{s_1s_2} y_{ks}
\]
where $s_1=\min\{s, j\}$, $s_2=\max\{s,j\}$, $\mathbf{Y}_{k[j]} = (Y_{k1},\dots, Y_{k(j-1)}, Y_{k(j+1)},\dots, Y_{km})$. The computationally expensive term $\Lambda$ disappears. For a concise representation, we collect the main effects into the vector $\boldsymbol\beta=(\beta_1,\dots,\beta_m)^{\top}$ and the interaction (edge) effects into the vector $\boldsymbol\theta = (\theta_{12},\theta_{13},\dots,\theta_{(m-1)m})^{\top}$. In literature, the node-wise PL of the $j$th node and the global PL are defined as 
\[
PL_j(\boldsymbol\beta,\boldsymbol\theta)=\prod_{k=1}^n\Pr\left(Y_{kj}=y_{kj}|\mathbf Y_{k[j]}= \mathbf y_{k[j]}\right) \quad \mbox{and} \quad PL(\boldsymbol\beta,\boldsymbol\theta) = \prod_{j=1}^m PL_j(\boldsymbol\beta,\boldsymbol\theta),
\]
respectively. Both of these PLs can be solved by software that solves GLMs. For edge selection, $l_1$ regularization (on $\theta_{j_1j_2}$'s) is generally applied to the PL, say PLL1. Node-wise PLL1  \citep{Ravikumar2010}, and global PLL1 \citep{Canditiis2020} are examples. \citet{Brusco2023} concluded that the node-wise PLL1 outperforms the global PLL1 under their simulation scenarios. In short, PLL1s suffice to construct sparsely linked undirected graphs. 

Having covered estimation and model selection, we now shift our focus to hypothesis testing. Given a network, we are interested in modeling the interaction (edge) effect $\theta_{j_1j_2}$ and then testing the existence of the interaction effect with hypotheses $H_0:\theta_{j_1j_2}=0$ vs $H_a:\theta_{j_1j_2} \neq 0$. To this end, a proper estimation of the standard error is essential. If the PL approximates the true likelihood well, maximizing the PL should result in consistent estimates with proper standard error estimates. Unfortunately, the PL, after taking the derivative, only serves as estimating equations, and the standard error estimate is drastically underestimated in our simulations, see Section~\ref{ssec:SimuII}.  We think of finding parameter estimators and their standard errors using the generalized estimating equation (GEE; \citealp{Liang1986}) approach. As demonstrated in Section~\ref{ssec:SimuII}, combining PL and GEE yielded prominent biases when non-diagonal working correlation structures are considered, but ignorable biases when using the independent working correlation. In other words, the choice of working correlations matters. These motivate us to dive deeper into the estimation and hypothesis testing issues of the QEBD.

The choice of working correlation structures has been discussed in several aspects. \citet{Pepe1994} assert that if $E(Y_{kj}|\mathbf x_{kj}) = E(Y_{kj}|\mathbf x_{kj}, \mathbf x_{kj^{\prime}}, j^{\prime}\neq j )$ is incorrect, independent working correlation remains the only viable working covariance where $\mathbf x_{kj}$ denotes the $p$-dimensional covariate vector with respect to the $k$th observation at time $j$. Similarly, \citet{Pan2000} considered the transition model with Markov property, say $E(Y_{kj}|Y_{k(j-1)}, \mathbf x_{kj})$, and elaborated on the bias in GEE with dependent working covariance for specific linear models. They also conclude that the diagonal working correlation is valid for consistent estimation in linear transition models. For correlated binary variables, \citet{Bible2019} defined two transition models with random effects to account for subject-specific heterogeneity. In their cases, for hypothesis testing, the unstructured working correlation is suggested for their first model, and the bootstrap approach is recommended for their second model. These results point out that, for consistent estimations, the choice of working correlation may not be arbitrary, particularly when the mean model contains past information. 

Building upon the above literature review, we identify a methodological gap in applying the GEE approach to the PL with the conditional mean model $E(Y_{kj}|\mathbf Y_{k[j]}, \mathbf x_{kj})$ under the GLM framework. In this work, we establish that a diagonal working correlation ensures estimation consistency, whereas alternative structures such as exchangeable or AR(1) correlations may fail to do so. We further clarify how to correctly estimate the parameters of the QEBD and its regression counterpart, the QELR, via PL-based GEE. The remainder of the paper is organized as follows. Section~\ref{sec:GEE} reviews the properties of GEE with marginal means and extends them to the conditional mean setting, where our main theoretical result is also presented. This section additionally demonstrates that the Markov model emerges as a special case of the conditional mean model. Section~\ref{sec:QEDR} develops the estimating functions for PLs associated with QEBD and QELR. Section~\ref{sec:simu} reports simulation studies for Markov models, QEBDs, and QELRs. Section~\ref{sec:casestudy} applies the proposed methodology to two datasets, the carcinogenic toxicity of chemicals and the constitutional court opinion writing among justices, before concluding the paper.

\section{Generalized Estimation Equations} \label{sec:GEE}

We first fix the notation. Following the convention, we denote capital letters as matrices, e.g., $B$ and $C$; bold-faced letters as vectors, e.g., $\mathbf{h}$ and $\mathbf{y}$; bold-faced capital letters as a vector consisting of random variables, say $\mathbf{Y}$.  Next, define $\mathbf y_{[j]}\in \mathbb R^{m-1}$ as the vector of $\mathbf y$ but its $j$th element is dropped and define $\mathbf{y}_{[j]}^c \in \mathbb{R}^{m}$ as the vector of $\mathbf{y}$ but substitutes $c$ to the $j$th element of $\mathbf{y}$. For example, if $\mathbf{y}=(y_1, y_2, y_3)^{\top}$ then $\mathbf{y}_{[2]}=(y_1, y_3)^{\top}$ and $\mathbf{y}^0_{[2]}=(y_1, 0, y_3)^{\top}$. Throughout this paper, let $\mathbf{e}_j$ be the $j$th column of the $m$-dimensional identity matrix for $j=1,\dots,m$. For an $n\times m$ matrix $B$, let $[B]_j$ be the $j$th column of $B$ and $[B]_{ij}$ be the $(i,j)$th element of $B$. Denote $C\otimes B$ as the Kronecker product of matrices $C$ and $B$. In particular, for an $C \in \mathbb{R}^{2\times 3}$, the Kronecker product of $C$ and $B$ is
\[
    C \otimes B = \left[ \begin{array}{ccc} 
    c_{11}B & c_{12}B & c_{13}B\\
    c_{21}B & c_{22}B & c_{23}B\\
    \end{array}\right] \in \mathbb{R}^{2n\times 3m}
\]
where $c_{ij}=[C]_{ij}$. Moreover, let $vec(\cdot)$ be an operator that vectorizes its argument into a vector, e.g., $vec(C) = (c_{11}, c_{21}, c_{12}, c_{22}, c_{13}, c_{23})^{\top}$.  For a regression problem, consider $n$ independent pairs $(\mathbf{Y}_k, X_k)$, $k=1,\dots,n$, where $\mathbf{Y}_k \in \mathbb{R}^m$ and $X_k \in \mathbb{R}^{m \times p}$. Define $\mathbf{x}_{kj}$ as the $j$th column of matrix $X_k$. Under the GLM framework, in the view of $\mathbf y_k$, we consider a \textit{conditional mean model} as
\begin{equation} \label{eq:geeCM}
    g(E(Y_{kj}|\mathbf Y_{k[j]}, \mathbf x_{kj})) = \boldsymbol \beta^{\top}\mathbf x_{kj} + \boldsymbol\gamma^{\top}W_{kj}\mathbf y^0_{k[j]}
\end{equation}
where $g$ is the canonical link function, $W_{kj} \in \mathbb R^{q \times m}$, observed constants, and $\boldsymbol \psi = (\boldsymbol \beta^{\top}, \boldsymbol \gamma^{\top})^{\top} \in \mathbb R^{p+q}$, unknown parameters. Define $\mu_{kj}=E(Y_{kj}|\mathbf Y_{k[j]}, \mathbf x_{kj})$ and $\nu_{kj} = \partial g(\mu_{ij})/\partial d\mu_{kj}$. Moreover, define $\boldsymbol \mu_k=(\mu_{k1},\dots,\mu_{km})^{\top}$ and $A_k$ as a diagonal matrix with $[A_k]_{jj} = \nu_{kj}$, $j=1,\dots,m$. Also, we define the \textit{marginal mean model} as (\ref{eq:geeCM}) with $\boldsymbol\gamma = \mathbf 0$, the expectation of $Y_{kj}$ is unaffected by $\mathbf Y_{k[j]}$. Note that when a model is defined as a marginal model, our unknown parameter $\boldsymbol \psi$ is merely $\boldsymbol \beta$.



\subsection{GEE with Marginal Means} \label{ssec:GEEMM}

The seminal paper \citet{Liang1986} defines the GEE approach for consistent parameter estimation with robust standard error estimates. Following (\ref{eq:geeCM}) with $\boldsymbol \gamma = \mathbf 0$, the GEE can be defined as
\[
    \mathbf{U}(\boldsymbol{\psi}; R) = \sum_{k=1}^n \frac{\partial \boldsymbol{\mu}_k}{\partial \boldsymbol{\psi}} V_k^{-1}(\mathbf{Y}_k - \boldsymbol{\mu}_k)
    =\sum_{k=1}^n \sum_{j=1}^m \frac{\partial \mu_{kj}}{\partial \boldsymbol{\psi}}(\mathbf{Y}_k - \boldsymbol{\mu}_k)^{\top}V_k^{-1}\mathbf{e}_j
\]
where $V_k=A_k^{1/2} \Rr A_k^{1/2}$ is the so-called working covariance, and $\Rr$ is the working correlation indexed by the parameter vector $\boldsymbol \rho \in \mathbb{R}^q$.  The GEE estimator, $\hat\bt$, which satisfies the equation $\mathbf U(\hat\bt; \Rr)= \mathbf 0$, is consistent to $\bt_0$, which assures $E(\mathbf U(\bt_0; \Rr))= \mathbf 0$. The variance of $\hat\bt$ has the sandwich form $B^{-1}(\bt_0; \Rr)M(\bt_0; \Rr)B^{-1}(\bt_0; \Rr)$ where 
\[
M(\bt; \Rr) = E(\mathbf U(\bt; \Rr) \mathbf U^{\top}(\bt; \Rr))
\quad \mbox{and} \quad
B(\bt; \Rr) = E\left( -\partial \mathbf U(\bt; \Rr)/\partial \bt\right). 
\]
An estimator for the ``meat" is $\hat M(\hat\psi; R)$ where
\[
    \hat M(\bt; \Rr) = \sum_{k=1}^n \frac{\partial \mu_k}{\partial \bt} V_k^{-1}(\mathbf Y_k- \boldsymbol\mu_k)(\mathbf Y_k-\boldsymbol \mu_k)^{\top}V_k^{-1}\left[ \frac{\partial \mu_k}{\partial \bt}\right]^{\top}
\]
and an estimator of the ``bum" $B(\bt; \Rr)$ is $\hat B(\hat\bt; \Rr)$ where
\[
    \hat B(\bt; \Rr) = \sum_{k=1}^n \frac{\partial \boldsymbol\mu_k}{\partial\bt}V_k^{-1}\left[ \frac{\partial \boldsymbol \mu_k}{\partial\bt}\right]^{\top}.
\]
Detailed estimation procedures for $\boldsymbol \rho$ are provided in \cite{Liang1986}, \cite{geepack2006}, and \cite{Myers2010}. Under mild regularity conditions, the corresponding estimator is consistent for any choice of the working correlation, \cite{Liang1986}. Moreover, the GEE estimator is most efficient if the working correlation is correctly specified.

Variable selection and the working covariance selection are critical issues in practice \cite{Pan2001,Pan2002}. When considering GEE with marginal models, by mimicking the AIC \citep{Akaike1973}, \citet{Pan2001} defined the QIC under the quasi-likelihoods \citep{McCullagh1983} framework. Define $Q(\bt)$ as the (log) quasi-likelihood function and set $\partial Q(\bt)/\partial\bt = \mathbf U(\bt; \Rr)$. The existence conditions for such $Q$ are addressed in McCullagh and Nelder (1983). When the working correlation structure is not independent, $Q$ is complicated, and the resulting Kullback-Liebler distance (approximation) between the true model and the working model is untenable. Pan (2001), therefore, assumes the independent working correlation and defines QIC$=-2 Q(\hat\bt) + 2 \mbox{trace}(\hat \Omega \hat J)$
where 
\[
    \Omega =-E\left(\frac{\partial^2Q(\bt)}{\partial\bt\partial\bt^{\top}}\right)
    \quad \mbox{and} \quad
    J = Cov\left(\hat\bt\right)
\]
and $\hat\Omega$ and $\hat J$ are their estimates, respectively. For a parametric model, we substitute the log-likelihood function for the $Q$ function, and hence, $\mbox{trace}(\Omega J)=\mbox{trace}(I_p)=p$. In this case, the QIC and AIC coincide.

\subsection{GEE with Conditional Means} \label{ssec:GEECM}
In this subsection, we consider GEE with conditional means defined in (\ref{eq:geeCM}) with $\boldsymbol \gamma \neq \mathbf 0$. With the conditional mean and a pre-specified working correlation $\Rr$, the estimating functions can be written as
\begin{equation}\label{eq:gee_main}
    \boldsymbol \varphi(\bt; \Rr) = \sum_{k=1}^n \left[ \begin{array}{ccc} 
        \mathbf x_{k1} & \dots & \mathbf x_{km}\\
        W_{k1} \mathbf Y_{k[1]}^{0} & \dots & W_{km} \mathbf Y_{k[m]}^{0}
    \end{array}\right] A_k V_k^{-1}(\mathbf Y_k-\boldsymbol\mu_k) 
    \equiv \sum_{k=1}^n \widetilde{W}_kA_kV_k^{-1}(\mathbf Y_k-\boldsymbol\mu_k)
\end{equation}
where $A_k$ is a diagonal matrix with $[A_k]_{jj}=\nu_{kj}=\partial g(\mu_{kj})/d\mu_{kj}$, and $V_k=A_k^{1/2}\Rr A_k^{1/2}$. Note that $\nu_{kj}$ depends on all or part of the vector $\mathbf Y_{k[j]}$. With these formulations, the major conclusion of this work is summarized in Theorem~\ref{th:gee} below.

\begin{theorem}\label{th:gee}
    Consider the estimating function defined in (\ref{eq:gee_main}). With arbitrary working correlation $\Rr$, 
    \[
        E(\boldsymbol \varphi(\bt;\Rr)) = \sum_{k=1}^n\sum_{j=1}^m \left[ \begin{array}{c} 
            \mathbf 0 \\
            W_{kj}C_{kj}V_k^{-1} \mathbf e_j
        \end{array}\right]
    \]
    where $C_{kj} = E\left\{ \mathbf Y_{k[j]}^{0}(\mathbf Y_k-\boldsymbol\mu_k)^{\top}\nu_{kj}\right\}$. 
    If $\Rr$ is diagonal, $E(\boldsymbol \varphi(\bt;\Rr)) = \mathbf 0$.
\end{theorem}
\noindent In other words, the estimating functions $\boldsymbol\varphi(\bt; \Rr)$ result in consistent estimates if the working covariance matrix is diagonal, and otherwise, consistency is not guaranteed because $E(\boldsymbol \varphi(\bt;\Rr)) \neq \mathbf 0$, \citet{Stefanski2002}. 

Next, we spare some space to address the relevance of the robust variance estimation and of the QIC for GEEs with conditional means. Let $\hat\bt$ be the root of the GEE with working correlation $I_m$. Since $I_m$ is diagonal, by Theorem~\ref{th:gee}, $E(\boldsymbol \varphi(\bt;I_m))=\mathbf 0$. In this sequel, we have
\[
    M(\bt;I_m)=Cov\left(\boldsymbol\varphi(\bt;I_m)\right) =  E\left( \sum_{k=1}^n \widetilde{W}_k(\mathbf Y_k-\boldsymbol\mu_k)(\mathbf Y_k-\boldsymbol\mu_k)^{\top}\widetilde{W}_k^{\top}\right).
\]
Consequently, $\hat M(\bt; I_m)=\sum_{k=1}^n\widetilde{W}_k(\mathbf Y_k-\boldsymbol\mu_k)(\mathbf Y_k-\boldsymbol\mu_k)^{\top}\widetilde{W}_k^{\top}$ is an unbiased estimator of $M(\bt;I_m)$. Similarly, because
\[
    B(\bt;I_m)= E\left(-\frac{\partial \boldsymbol\varphi(\bt;I_m)}{\partial \bt}\right)= E\left(\sum_{k=1}^n \widetilde{W}_kA_k\widetilde{W}_k^{\top}\right), 
\]
we conclude that $\hat B(\bt;I_m)=\sum_{k=1}^n \widetilde{W}_kA_k\widetilde{W}_k^{\top}$
is an unbiased estimator for $B(\bt;I_m)$. Together, the sandwich formula is a proper estimator of the variance of $\hat\bt$. As for the relevance of using QIC for model selection, by substituting $\boldsymbol \varphi(\bt;I_m)$ to $\partial Q(\bt)/\partial \bt$, we conclude that $J$ can be consistently estimated by the sandwich estimates $\hat B^{-1}(\hat\bt; I_m)\hat M(\hat\bt; I_m)\hat B^{-1}(\hat\bt; I_m)$ and $\Omega$ can be estimated unbiasedly by $\hat B(\hat\bt; I_m)$. Therefore, $\mbox{trace}(\hat\Omega\hat J) = \mbox{trace}\left(\hat M(\hat\bt; I_m) \hat B^{-1}(\hat\bt; I_m)\right)$.
For conditional mean models, since both the meat matrix and the bum matrix are consistent estimates, using QIC for model selection is relevant.

In short, when considering GEE with conditional mean models, the independent working correlation guarantees estimation consistency while other working correlations do not. Moreover, the robust standard error estimates and the model selection criterion QIC is still valid.

\subsection{Markov Model as an Example} \label{ssec:TM}
At time $t$, define the collection of the past information as $\mathcal{H}_{kt} =\{\mathbf x_{kt}, (y_{k(t-1)}, \mathbf x_{k(t-1)}), \dots,(y_{k1}, \mathbf x_{k1})\}$ and $\mathcal{H}_{k1}=\{\mathbf x_{k1}\}$. The transition model defines conditional density functions $f(y_{kt}| \mathcal{H}_{kt})$, $t=1,2,\dots, m$, so the joint density function is
\[
    f(\mathbf y_k)=f(y_{km}|\mathcal{H}_{km})f(y_{k(m-1)}|\mathcal{H}_{k(m-1)}) \times \cdots \times f(y_{k1}|\mathcal{H}_{k1}) = \prod_{t=1}^m f(y_{kt}|\mathcal{H}_{kt}).
\]
For longitudinal data, the mean response of $Y_{kt}$ can be modeled with their previous observations, $(y_{k1}, y_{k2}, \dots, y_{k(t-1)})$. Define the conditional mean function $\pi_{kt} = \Pr(Y_{kt}=1|Y_{k(t-1)}=y_{k(t-1)},\dots,Y_{k1}=y_{k1})$. The $q$th-order Markov logistic regression model has the form $\mbox{logit}\left(\pi_{kt}\right) = \mathbf{x}_{kt}^{\top}\boldsymbol{\beta} + \sum_{s=1}^q\gamma_{s} y_{k(t-s)}$, $t=1,\dots,m$, where $\Pr(Y_{k(t-s)}=0)=1$ for $t-s\leq 0$. The joint distribution of $\mathbf Y_k$ is $\Pr(\mathbf Y_k = \mathbf y_k)=\prod_{t=1}^m \pi_{t}^{y_{kt}}(1-\pi_{kt})^{1-y_{kt}}$ which has exactly the same form as a logistic regression. The corresponding score function is $\sum_{t=1}^m \left[ \mathbf x_{kt}^{\top}, Y_{k(t-1)},\dots,Y_{k(t-q)}\right]^{\top} (Y_{kt}-\pi_{kt})$.
Consequently, the Markov logistic regression models have the form of the conditional mean model. Thus, Theorem~\ref{th:gee} applies.

\section{Quadratic Exponential Distributions and Regressions} \label{sec:QEDR}

\subsection{Quadratic Exponential Binary Distributions}

First, we rewrite (\ref{eq:Asy_Ising}) in a quadratic form analogous to the normal distribution. 
Let $\Theta$ be an $m\times m$ symmetric matrix such that $[\Theta]_{jj}=0$ and $[\Theta]_{j_1j_2} =  \theta_{j_1j_2}=[\Theta]_{j_2j_1}$ for $j_1<j_2$. By collecting all unique parameters in $\Theta$ into $\boldsymbol{\theta}=(\theta_{12},\dots,\theta_{1m}, \theta_{23}, \dots,\theta_{(m-1)m})^{\top} \in \mathbb{R}^{(m-1)m/2}$, the unknown parameter vectors of QEBD are $\bt = (\boldsymbol{\beta}^{\top}, \boldsymbol{\theta}^{\top})^{\top}$. Then
\[
    \Pr(\mathbf Y_k =\mathbf y_k)= \exp\left\{ \mathbf y_k^{\top}\boldsymbol{\beta}+ \frac{1}{2}\mathbf y_k^{\top} \Theta \mathbf y_k - \Lambda\right\}.
\]
Solving the MLE of $\bt$ via the Newton algorithm seems to be feasible. However, it requires evaluating $\Lambda$ repeatedly, and hence, finding MLE causes a massive computation burden, even when $m$ is mild, say $m=15$.

Following the PL approach \citep{Strauss1990}, in particular, we posit the conditional probability $\pi_{kj}=\Pr(Y_{kj}=1|\mathbf{Y}_{k[j]}=\mathbf{y}_{k[j]})$ with a logistic regression form
\begin{equation} \label{eq:cIsing}
    \mbox{logit}(\pi_{kj})= \beta_j + \sum_{s\neq j} [\Theta]_{sj}y_{ks}
    = \mathbf e_j^{\top} \boldsymbol \beta + \mathbf{e}_j^{\top} \Theta\mathbf{y}_{k[j]}^0
\end{equation}
for $j=1,\dots,m$. After some manipulation, the estimating functions of QEBD become
\begin{equation} \label{eq:EE_QEBD}
    \boldsymbol \varphi(\bt, \Rr) = \sum_{k=1}^n \left[ \begin{array}{ccc} \mathbf e_1 & \dots & \mathbf e_m \\ G(I_m \otimes \mathbf e_1)\mathbf Y_{k[1]}^0 & \dots & G(I_m\otimes \mathbf e_m)\mathbf Y_{k[m]}^0 \end{array}\right]A_kV_k^{-1}(\mathbf Y_k - \boldsymbol \pi_k).
\end{equation}
%
%
The definition of $G$ is as below. Define $\mathbf{g}_{ij} =  (\mathbf{e}_i\otimes \mathbf{e}_j + \mathbf{e}_j\otimes\mathbf{e}_i)$ and 
\[
    G^{\top} = \left[\mathbf{g}_{12}, \mathbf{g}_{13},\dots,\mathbf{g}_{1m}, \mathbf{g}_{23},\mathbf{g}_{24},\dots, \mathbf{g}_{(m-1)m}\right] \in \mathbb{R}^{m^2 \times m(m-1)/2}.
\]
Thus, $vec(\Theta)=G^{\top} \boldsymbol \theta$. 
%
%
Because $\mathbf Y_{[j]}^0 \otimes \mathbf e_j = (I_m \otimes \mathbf e_j)\mathbf Y_{[j]}^0$, the last component of (\ref{eq:cIsing}) can be rewritten as
$
    \mathbf e_j^{\top}\Theta \mathbf Y_{[j]}^0 = (\mathbf Y_{[j]}^0 \otimes \mathbf e_j)^{\top}vec(\Theta)=\left\{(I_m \otimes \mathbf e_j)\mathbf Y_{[j]}^0\right\}^{\top}G^{\top}\boldsymbol \theta.
$
This suffices to result in (\ref{eq:EE_QEBD}).

Consequently, the estimation functions for $\bt$ are equivalent to those in (\ref{eq:gee_main}) with $\mathbf x_{kj}=\mathbf e_j$ and $W_{kj}= G(I_m\otimes \mathbf e_j)$. According to Theorem~\ref{th:gee},  under an arbitrary working correlation, the expectation of (\ref{eq:EE_QEBD}) does not necessarily vanish as the sample size goes to infinity. Hence, solving $\boldsymbol \varphi(\boldsymbol{\bt};\Rr)=\mathbf{0}$ may yield biased estimations except for some carefully chosen working correlation $\Rr$. 

\subsection{Quadratic Exponential Logistic Regressions}

In this subsection, we suppress the subscript $k$ to have a clearer representation. Further, we name $\beta_j$'s as main effects and $\theta_{j_1j_2}$'s as interaction effects. Constructing models for the main effect is relatively simple. Suppose $\mathbf{x}_j$ is a vector of predictors affecting the individual effect. Then, we can define $\beta_j = \tilde{\boldsymbol{\beta}}^{\top}\mathbf{x}_j$. Modeling the main effects in this way has been proposed in the literature \citep{Connolly1988, Zhao1990, Joe1996}. Next, we deal with the interaction effect, $\theta_{j_1j_2}$'s. As demonstrated in \citet{Arnold1989}, arbitrarily modifying the fully conditional distributions may not yield a unique joint distribution. The compatibility condition is required to ensure the existence and uniqueness of the joint distribution. The compatibility condition for the model (\ref{eq:cIsing}) is that $\Theta$ is symmetric, \citet{Joe1996}. Consequently, when having extra information about the interaction effects, we may define $[\Theta]_{j_1j_2} = \gamma w_{j_1j_2}$ where $w_{j_1j_2}$ is a predictor representing the cause of the interaction between the $j_1$th variable and the $j_2$th variable, and $\gamma$ is the corresponding regression coefficient. Additionally, we need to enforce that $\gamma w_{j_1j_2}=\gamma w_{j_2j_1}$ to satisfy the compatible condition. 


When there are $L$ characteristics that potentially describes the interactions among variable $y_j$'s, to meet the compatibility condition, we  define $w^{\ell}_{ij} = \mbox{dist}(\mathbf{u}^{\ell}_i, \mathbf{u}^{\ell}_j)$ where $\mathbf{u}^{\ell}_j$ is the observed vector of the $\ell$th characteristic about the $j$th variable, and a distance function $\mbox{dist}(\mathbf{u}^{\ell}_i, \mathbf{u}^{\ell}_j) \geq 0$ which returns the distance between two vectors where $\mbox{dist}(\mathbf{u}^{\ell}_i, \mathbf{u}^{\ell}_j) = 0$ if and only if $\mathbf{u}^{\ell}_i=\mathbf{u}^{\ell}_j$ and the distance function has to be symmetric, $\mbox{dist}(\mathbf{u}^{\ell}_i, \mathbf{u}^{\ell}_j)=\mbox{dist}(\mathbf{u}^{\ell}_j, \mathbf{u}^{\ell}_i)$. 
The proposed quadratic exponential logistic regression (QELR) is, therefore, for $j=1,\dots,m$, having the fully conditional log-density function
\begin{equation}\label{eq:fcIsing}
\mbox{logit}\left(\Pr(Y_j=1|\mathbf{Y}_{[j]} =\mathbf{y}_{[j]})\right) \\
    = \tilde{\boldsymbol{\beta}}^{\top}\mathbf{x}_j + \sum_{i\neq j} \sum_{\ell=1}^L \gamma^{\ell}w_{ij}^{\ell} y_i
    = \tilde{\boldsymbol{\beta}}^{\top}\mathbf{x}_j + \boldsymbol{\gamma}^{\top} W_j \mathbf{y}^0_{[j]}
\end{equation}
where $\boldsymbol{\gamma}=(\gamma_1,\dots,\gamma_L)^{\top}$ and
\[
    W_j = \left[ \begin{array}{cccc}
    w^1_{1j} & w^1_{2j} & \dots & w^1_{mj}\\
    w^2_{1j} & w^2_{2j} & \dots & w^2_{mj}\\
    \vdots & \vdots & \ddots & \vdots\\
    w^L_{1j} & w^L_{2j} & \dots & w^L_{mj}\\
    \end{array}\right] \in \mathbb{R}^{L\times m}.
\]
The simplest version of (\ref{eq:fcIsing}) is setting a common interaction effect, shorthanded as QELR-CI. That is, 
\begin{equation} \label{eq:simplest}
    \mbox{logit}\left(\Pr(Y_j=1|\mathbf{Y}_{[j]}=\mathbf{y}_{[j]})\right) 
    =\beta_j + \sum_{s\neq j} \theta y_s
    = \beta_j + \theta \left(\sum_{s=1}^m y_s - y_j\right)
\end{equation}
for $j=1,\dots, n$. In contrast to the interaction effects in (\ref{eq:fcIsing}), this model has $\gamma = \theta \quad \mbox{and} \quad W_j = [1,\dots,1] \in \mathbb{R}^{1\times m}$. This equation implies the symmetry among variable in $\mathbf{y}_{[i]}$ and thus $\sum_{i=1}^m y_i$ is a sufficient statistic for $\theta$, \citet{Connolly1988}. \citet{Qu1987} and \citet{Connolly1988} considered a more general model, $\beta_j + F_{\boldsymbol{\alpha}}(\sum_{s=1}^m y_s - y_j)$ where $F_{\boldsymbol{\alpha}}$ is a known function with unknown parameters $\boldsymbol{\alpha}$.


With the compatible condition,  defining all fully conditional distributions like (\ref{eq:fcIsing}) results in a unique joint model. Solving the joint model likelihood is challenging. The difficulty arises due to evaluating the normalizing term, which consists of $2^m$ terms per observation. According to the previous discussion and our simulation results listed in Simulation Studies, we suggest using the GEE approach. 
Note that the upper half of the estimating functions in (\ref{eq:gee_main}) has the form $[\mathbf{x}_1,\dots,\mathbf{x}_m]AV^{-1}(\mathbf{Y}-\boldsymbol{\pi})$ which has mean zero. On the other hand, the lower half of the estimating functions has a complicated form of $y_i$'s, which may not yield zero means. Hence, the consistency of the estimation is questionable. Again, arbitrary choices of the working covariance are not guaranteed to end up with consistent estimates. According to Theorem~\ref{th:gee}, the estimating equations (\ref{eq:gee_main}) result in a biased estimation in general, except that $V$ is diagonal.

\section{Simulations}\label{sec:simu}

In the simulation study, we compare the following estimation approaches. The first is the maximum likelihood estimator (MLE), obtained by directly maximizing the likelihood function (\ref{eq:Asy_Ising}). The second approach is based on the global PL (GPL). We implement GPL in two ways. First, treat the GPL as a GLM likelihood and apply standard GLM software. Hereafter, we shorten this approach as GGLM. Second, solve GPL via GEE with three working correlation structures—independence (GEE-IND), exchangeable (GEE-EXC), and AR(1) (GEE-AR1). These GEE-based methods are implemented using the \textit{geepack} package \citep{geepack2006} in R \citep{R2021}. From the asymptotic properties of these estimators, we anticipate the following: the point estimates from GGLM and GEE-IND coincide, although their standard error estimates differ. In particular, when the data are generated from a QEBD, the standard errors from GLM with GPL likelihood are invalid. For GEE-EXC and GEE-AR1, Theorem \ref{th:gee} implies that the resulting estimators may be inconsistent. Overall, we expect the performance of GEE-IND to be comparable to that of MLE in applicable scenarios. R codes for all simulation studies are available at \url{https://github.com/jonong03/QELR/}.

We evaluate the performance of the aforementioned approaches in the following metrics. Bias is the average of parameter estimates subtracting the true value. When the MLE is tenable ($m<15$), S.E. represents the average of standard error estimates, and R.E. is the S.E. of one particular method divided by the S.E. of the MLE. When obtaining MLE is inefficient ($m \geq 15$), the Emp. S.D., the sample standard deviation of the GGLM estimates, is listed. The subsequent R.E. is therefore the S.E. of one particular method divided by the Emp. S.D..

\subsection{Simulation I: Transition Model}

We illustrate the implications of Theorem~\ref{th:gee} using the Child’s Respiratory Illness data \citep{Agresti2019}. In this dataset, children were evaluated annually for the presence of respiratory illness at ages 7 through 10, with maternal smoking status as a key covariate. We adopt the following first-order Markov model for the conditional mean:
\begin{equation} \label{eq:FirstOrderMarkov}
    \mbox{logit}(\Pr\left(Y_{t}=1|Y_{t-1}=y_{t-1})\right)
    = \beta_0 + \beta_1 S + \beta_2 t + \gamma_1 y_{t-1}, 
\end{equation}
$t=8,9, 10$, and $\mbox{logit}(\Pr\left(Y_{7}=1)\right) = \beta_0 + \beta_1 S + \beta_2 \times 7$, where  $S=1$ if smoking regularly and $S=0$ otherwise. These conditional probabilities assemble the joint likelihoods \citep{Diggle2002} and hence the GPL is exactly the same as the likelihood of the first-order Markov model. For demonstration, we simulated the data following the above assumptions and then applied the aforementioned approaches. Table~\ref{tb:Markov} summarizes the estimation results. The significant discrepancies between the GGLM and GEE-AR1's standard error estimates and between the GEE-IND and GEE-EXC's (and GEE-AR1's) regression coefficient estimates motivate us to look into the theoretical properties of conditional mean models.  

The estimation results, summarized in Table~\ref{tb:Markov}, highlight substantial differences across methods. As the MLE can be computed via GGLM, its bias and standard error serve as benchmarks. For GEE-IND, the estimated biases closely match those from PL, while the standard errors are slightly smaller, with relative efficiencies between 0.970 and 0.989. For GEE-EXC, the regression coefficient estimates remain nearly unbiased, but the standard error of $\hat\gamma$ (the coefficient for $y_{t-1}$) is drastically underestimated. In contrast, GEE-AR1 exhibits pronounced biases for both $\hat\beta_0$ and $\hat\gamma$, and hence its standard error estimates are questionable. Overall, these results advocate Theorem~\ref{th:gee}: when fitting conditional mean models, only GEE-IND guarantees consistent estimation.

\begin{table}
\begin{center}
  \caption{Simulation Results of the First-Order Markov Logistic Regression with $n=300$ and replicates$=500$. The Bias columns show the true value subtracted from the averages of 500 estimates. The S.E. column for MLE shows the average of 500 standard error estimates. Each of the R.E. columns for GEEs is the average of 500 standard error estimates divided by the corresponding S.E. of MLE. \label{tb:Markov}}
  \begin{tabular}{@{\extracolsep\fill}crrrrrrrrrr@{\extracolsep\fill}}
    \hline
        & & \multicolumn{2}{c}{MLE} & \multicolumn{2}{c}{GEE-IND} & \multicolumn{2}{c}{GEE-EXC} & \multicolumn{2}{c}{GEE-AR1}\\
         \cmidrule(l){3-4} \cmidrule(l){5-6} \cmidrule(l){7-8}\cmidrule(l){9-10}
        Variable & Truth & Bias & S.E. & Bias & R.E. & Bias & R.E. & Bias & R.E.\\
        \hline
  $\beta_0$ & 0.423 & -0.001 & 0.707 & -0.001 & 0.989 & 0.001 & 0.969 & \textbf{-0.060} & \textbf{0.898} \\ 
  $\beta_1$ & 0.223 & -0.004 & 0.190 & -0.004 & 0.970 & -0.005 & 0.973 & 0.001 & 0.965 \\ 
  $\beta_2$ & -0.316 & -0.001 & 0.086 & -0.001 & 0.988 & -0.002 & 0.952 & 0.008 & \textbf{0.851} \\ 
  $\gamma$ & 2.180 & -0.007 & 0.213 & -0.007 & 0.989 & 0.001 & \textbf{0.551} & \textbf{-0.121} & \textbf{0.354} \\ 
\hline
\end{tabular} \end{center}
\end{table}

\subsection{Simulation II: Quadratic Exponential Distributions} \label{ssec:SimuII}
The second simulation study assessed both the estimation accuracy and computational efficiency of MLE, GPL, and GEE-based methods. Data were generated from the QEBD model (\ref{eq:Asy_Ising}) with $m$ binary responses and a fixed sample size of $n=300$, under prespecified parameter values. For each scenario, parameters were estimated using MLE, GGLM, and GEE-IND; GEE-EXC was excluded because it failed to converge in this setup, frequently. A total of 500 datasets were simulated, and the results are summarized in Table~\ref{tb:QED}. We observe that both MLE and GEE-IND yielded negligible biases and almost identical standard error estimates on average. However, although the GGLM approach yields the same estimates as the GEE-IND, its standard error estimates were far too small. We conclude that the MLE and GEE-IND are numerically comparable. 

To evaluate computational burden, we fixed $n=300$ and varied the number of binary responses $m$. The average computing times for MLE were 0.652, 26.346, and 196.759 seconds for $m=5, 10,$ and $12$, respectively, compared with 0.064, 0.213, and 0.320 seconds for GEE-IND. These results highlight the steep computational cost of MLE as the dimension increases, rendering it impractical for larger models (e.g., $m\geq 15$). In contrast, the GEE-IND approach scales efficiently and provides a practical alternative for high-dimensional binary data.

\begin{table}
\begin{center}
  \caption{Parameter Estimations for the QEBD with $n=300$, $m=5$, and  replicates$=500$. The ``Emp. S.D." is the sample standard deviation of the 500 MLE estimates. The Bias columns show the true value subtracted from the averages of 500 estimates. Each of the R.E. columns is the average of 500 standard error estimates divided by the corresponding ``Emp. S.D." Each of the PW columns shows the rejection rates over the 500 replications under the null hypothesis that the parameter is equal to zero.}
\label{tb:QED}
  \begin{tabular}{@{\extracolsep\fill}crrrrrrrrrrrc@{\extracolsep\fill}}
    \hline
         ~ & ~ & Emp. & \multicolumn{3}{c}{MLE} & \multicolumn{3}{c}{GGLM} & \multicolumn{3}{c}{GEE-IND} \\
         \cmidrule(l){4-6} \cmidrule(l){7-9} \cmidrule(l){10-12}
        Variable & Truth & S.D. & Bias & R.E. & PW & Bias & R.E.$^1$ & PW & Bias & R.E. & PW \\
        \hline
            $\beta_1$     & -1.500 & 0.318 & -0.031 & 1.012 & 0.998 & -0.032 & 0.778 & 1.000 & -0.032 & 1.012 & 1.000 \\ 
            $\beta_2$     & -0.750 & 0.288 & 0.002 & 0.969 & 0.756 & 0.001 & 0.665 & 0.914 & 0.001 & 0.968 & 0.758 \\
            $\beta_3$     & 0.000 & 0.272 & 0.010 & 1.022 & 0.044 & 0.010 & 0.699 & 0.178 & 0.010 & 1.022 & 0.046 \\
            $\beta_4$     & 0.750 & 0.283 & 0.010 & 0.966 & 0.792 & 0.011 & 0.670 & 0.914 & 0.011 & 0.966 & 0.792 \\ 
            $\beta_5$     & 1.500 & 0.280 & 0.011 & 0.991 & 1.000 & 0.012 & 0.694 & 1.000 & 0.012 & 0.991 & 1.000 \\
            $\theta_{12}$ & -0.400 & 0.262 & 0.007 & 1.007 & 0.328 & 0.007 & 0.804 & 0.472 & 0.007 & 1.006 & 0.328 \\ 
            $\theta_{13}$ & 1.200 & 0.243 & 0.024 & 1.033 & 0.996 & 0.024 & 0.725 & 1.000 & 0.024 & 1.033 & 0.996 \\
            $\theta_{14}$ & 0.000 & 0.269 & -0.007 & 0.947 & 0.056 & -0.007 & 0.656 & 0.192 & -0.007 & 0.946 & 0.056 \\ 
            $\theta_{15}$ & 0.000 & 0.266 & -0.004 & 0.995 & 0.058 & -0.004 & 0.683 & 0.162 & -0.004 & 0.995 & 0.058 \\ 
            $\theta_{23}$ & -0.400 & 0.263 & -0.029 & 1.003 & 0.370 & -0.028 & 0.794 & 0.532 & -0.028 & 1.003 & 0.366 \\ 
            $\theta_{24}$ & 0.000 & 0.253 & 0.016 & 0.996 & 0.054 & 0.016 & 0.692 & 0.166 & 0.016 & 0.995 & 0.056 \\ 
            $\theta_{25}$ & 0.000 & 0.269 & -0.008 & 0.975 & 0.048 & -0.008 & 0.671 & 0.188 & -0.008 & 0.975 & 0.050 \\
            $\theta_{34}$ & 0.000 & 0.251 & -0.001 & 0.988 & 0.056 & -0.002 & 0.804 & 0.128 & -0.002 & 0.989 & 0.060 \\ 
            $\theta_{35}$ & 0.000 & 0.259 & -0.020 & 0.985 & 0.066 & -0.020 & 0.683 & 0.178 & -0.020 & 0.985 & 0.066 \\
            $\theta_{45}$ & -0.400 & 0.254 & 0.023 & 1.042 & 0.284 & 0.023 & 0.829 & 0.442 & 0.023 & 1.042 & 0.290 \\ 
                \hline
\end{tabular}\\
$^1$: The R.E.s are far away from 1 in this column, which means that the standard errors provided by GGLM are too small.
\end{center}
\end{table}

\subsection{Simulation III: Quadratic Exponential Logistic Regressions}

Next, we present a simulation study to assess the performance of our proposed QELRs with common interaction (\ref{eq:simplest}) and linear interaction (\ref{eq:fcIsing}). We simulated 500 datasets, each consisting of $n$ individuals and $m=15$ correlated binary responses. The sample sizes $n$ are 100, 300, and 500. In particular, the conditional mean model of the QELR-CI is
\[
    \mbox{logit}\left(\Pr(Y_j=1|\mathbf{Y}_{[j]}=\mathbf{y}_{[j]})\right)
    = \beta_0 + \beta_1 x_{1j} + \beta_2 x_{2j} + \gamma \left(\sum_{i=1}^m y_i - y_j\right)
\]
and the conditional mean model of the QELR with linear interactions is
\[
    \mbox{logit}\left(\Pr(Y_j=1|\mathbf{Y}_{[j]}=\mathbf{y}_{[j]})\right)
    = \beta_0 + \beta_1 x_{1j} + \beta_2 x_{2j} + \sum_{i=1, i\neq j}^m (\gamma_1 w_{ij}^1+\gamma_2w_{ij}^2)y_i.
\]
Note that $w_{ij}^{\ell}$ is a similarity measure of the $i$th and the $j$th variables. The variables $u_i^{\ell}$s were sampled from the set $\{1, 2, 3\}$ uniformly and independently so that $\Pr(w_{ij}^{\ell}=1)=\Pr(u_i^{\ell}=u_j^{\ell})=1/3$. The compatibility condition of \cite{Joe1996} is satisfied by doing so.


We compared the GGLM, GEE-IND, and GEE-EXC approaches. As shown in Tables~\ref{tb:QELR_C} and \ref{tb:QELR_L}, both GGLM and GEE-IND produced negligible biases, but GGLM consistently underestimated standard errors, leading to inflated significance of hypothesis tests. GEE-EXC showed substantial biases and a 57\% divergence rate under the common interaction model, though performance improved under the linear interaction model, with negligible biases and no divergence. These results confirm that GEE-IND provides consistent estimation, while other working correlations can yield biased or unstable results.

\begin{table}
\begin{center}
  \caption{Estimation Results of the QELR-CI Model with $m=15$. The ``Emp. S.D." is the sample standard deviation of the 500 GGLM estimates. The Bias columns show the true value subtracted from the averages of 500 estimates. Each of the R.E. columns is the average of 500 standard error estimates divided by the corresponding ``Emp. S.D." }
\label{tb:QELR_C}
  \begin{tabular}{@{\extracolsep\fill}ccrrrrrrrr@{\extracolsep\fill}}
        \toprule
        & & True & Emp. & \multicolumn{2}{c}{GGLM} & \multicolumn{2}{c}{GEE-IND} &\multicolumn{2}{c}{GEE-EXC*}\\
        \cmidrule(l){5-6} \cmidrule(l){7-8} \cmidrule(l){9-10}
        n & Parameter & Value & S.D. & Bias & R.E.$^1$ & Bias & R.E. & Bias$^2$ & R.E. \\
        \midrule
        100 & $\beta_{0}$  & -2.4 &  0.594 & 0.042 & 0.627 & 0.042 & 0.975 & 0.918 & 0.976 \\ 
          & $\beta_{1}$      & -2.0 &  0.220 & -0.051 & 0.978 & -0.051 & 0.982 & 0.161 & 0.886 \\ 
          & $\beta_{2}$      & -2.6 &  0.265 & -0.068 & 0.948 & -0.068 & 0.957 & 0.216 & 0.867 \\  
          & $\gamma$         & -1.4 &  0.298 & -0.082 & 0.555 & -0.082 & 0.937 & -0.438 & 0.999 \\ 
        \midrule
        300 & $\beta_{0}$ & -2.4 &  0.333 & 0.014 & 0.623 & 0.014 & 0.965 & -0.863 & 0.943 \\ 
         & $\beta_{1}$      & -2.0 &  0.119 & -0.015 & 1.019 & -0.015 & 1.041 & 0.638 & 0.929 \\  
         & $\beta_{2}$      & -2.6 &  0.142 & -0.020 & 0.992 & -0.020 & 1.018 & -0.301 & 0.924 \\  
         & $\gamma$         & -1.4 &  0.167 & -0.023 & 0.546 & -0.023 & 0.938 & 0.969 & 1.026 \\ 
        \midrule
        500 & $\beta_{0}$ & -2.4 & 0.247 & 0.003 & 0.649 & 0.003 & 1.000 & -0.127 & 0.991 \\ 
          & $\beta_{1}$     & -2.0 & 0.090 & -0.011 & 1.034 & -0.011 & 1.059 & -1.248 & 0.941 \\  
          & $\beta_{2}$     & -2.6 & 0.104 & -0.011 & 1.041 & -0.011 & 1.074 & 0.127 & 0.969 \\  
          & $\gamma$        & -1.4 & 0.125 & -0.013 & 0.560 & -0.013 & 0.963 & 0.371 & 1.061 \\  
\hline
\end{tabular}\\
*: Among the 500 replications, 57\% of them resulted in divergence estimations and were excluded from calculating the averages and standard errors.\\
$^1$: Some of the R.E. in this column are prominent, especially for $\beta_0$ and $\gamma$ estimates, which means that the standard error estimates of the GGLM are too small.\\
$^2$: The biases in this column are prominent, which means that the estimators of GEE-EXC are biased.
\end{center}
\end{table}

\begin{table}
\begin{center}
  \caption{Estimation Results of the QELR with Linear Interaction Effects with $m$=15. The ``Emp. S.D." is the sample standard deviation of the 500 GGLM estimates. The Bias columns show the true value subtracted from the averages of 500 estimates. Each of the R.E. columns is the average of 500 standard error estimates divided by the corresponding ``Emp. S.D."}
\label{tb:QELR_L}
  \begin{tabular}{@{\extracolsep\fill}ccrrrrrrrr@{\extracolsep\fill}}
        \toprule
        & & True & Emp. & \multicolumn{2}{c}{GGLM} & \multicolumn{2}{c}{GEE-IND} &\multicolumn{2}{c}{GEE-EXC}\\
        \cmidrule(l){5-6} \cmidrule(l){7-8} \cmidrule(l){9-10}
        n & Parameter & Value & S.D. & Bias & R.E.$^1$ & Bias & R.E. & Bias$^2$ & R.E. \\
        \midrule
         100 & $\beta_{0}$ & -2.4 & 0.242 & -0.008 & 0.882 & -0.008 & 1.000 & 0.127 & 0.997 \\ 
        & $\beta_{1}$      & -2.0   & 0.145 & -0.035 & 1.083 & -0.035 & 1.084 & -0.018 & 1.076 \\
        & $\beta_{2}$      & -2.6   & 0.178 & -0.031 & 1.032 & -0.031 & 1.037 & -0.009 & 1.032 \\
        & $\gamma_{1}$     & -1.4   & 0.238 & -0.045 & 0.761 & -0.045 & 0.976 & -0.123 & 0.990 \\ 
        & $\gamma_{2}$     & -0.5   & 0.193 & -0.023 & 0.807 & -0.023 & 1.011 & -0.095 & 1.028 \\
        \midrule
        300 & $\beta_{0}$ & -2.4 & 0.142 & 0.002 & 0.860 & 0.002 & 0.981 & 0.135 & 0.976 \\ 
        & $\beta_{1}$       & -2.0 & 0.087 & -0.012 & 1.030 & -0.012 & 1.038 & 0.004 & 1.031 \\  
        & $\beta_{2}$       & -2.6 & 0.104 & -0.011 & 1.009 & -0.011 & 1.018 & 0.009 & 1.012 \\  
        & $\gamma_{1}$      & -1.4 & 0.136 & -0.018 & 0.757 & -0.018 & 0.989 & -0.097 & 1.005 \\  
        & $\gamma_{2}$      & -0.5 & 0.111 & -0.008 & 0.802 & -0.008 & 1.030 & -0.082 & 1.049 \\ 
        \midrule
        500 & $\beta_{0}$ & -2.4 &  0.113 & 0.003 & 0.833 & 0.003 & 0.953 & 0.135 & 0.948 \\  
        & $\beta_{1}$       & -2.0 &  0.067 & -0.008 & 1.026 & -0.008 & 1.033 & 0.008 & 1.027 \\  
        & $\beta_{2}$       & -2.6 &  0.082 & -0.010 & 0.992 & -0.010 & 1.003 & 0.010 & 0.997 \\  
        & $\gamma_{1}$      & -1.4 &  0.108 & -0.009 & 0.738 & -0.009 & 0.973 & -0.088 & 0.990 \\  
        & $\gamma_{2}$      & -0.5 &  0.089 & -0.011 & 0.771 & -0.011 & 0.988 & -0.084 & 1.007 \\  
\hline
\end{tabular}\\
$^1$: Some of the R.E. in this column are prominent, especially for $\beta_0$ and interaction effects $\gamma_1$ and $\gamma_2$ estimates, which means that the standard error estimates of the GGLM are too small.\\
$^2$: Some of the biases in this column are prominent, especially for $\beta_0$ and interaction effects $\gamma_1$ and $\gamma_2$ estimates, which means that the estimators of GEE-EXC are biased.
\end{center}
\end{table}

\section{Case Studies} \label{sec:casestudy}

\subsection{Carcinogenic Toxicity of Chemicals}

\citet{Haseman1990} discussed four \textit{in vitro} assays for genetic toxicity, which were investigated for their ability to predict the carcinogenicity of chemicals. These assays were mutagenesis in Salmonella typhimurium (SAL), mouse lymphoma cells (MLA), chromosome aberrations (ABS), and sister chromatid exchanges in Chinese hamster ovary cells (SCE). Each of 95 selected chemicals was individually examined to determine whether it is carcinogenic in the four assays mentioned above. For each chemical, a 4-tuple of binary responses (1: carcinogenic and 0: non-carcinogenic) was recorded. The data is listed in Table~1 of \citet{Lipsitz1994}. 

In addition to parameter estimation, we performed edge/interaction effect selection using a backward elimination procedure based on QIC. Starting from the full model, interaction terms were sequentially removed whenever their exclusion improved/lowered the QIC, and the process continued until no further improvement was possible. Table~\ref{tb:CTC} summarizes the results. The QEBD$_F$ columns report GEE-IND estimates from the full model, whereas the QEBD$_R$ columns present estimates after backward elimination, in which two interaction effects (SAl–SCE and MLA–ABS) were removed. In both models, all retained interaction effects were non-negative, consistent with the expectation that a chemical identified as carcinogenic by one method is more likely to be identified by others. We also applied the QELR-CI, which yielded a positive common interaction estimate ($\hat\gamma=1.566$). However, among the three models, QEBD$_R$ achieved the smallest QIC, indicating that QELR-CI is overly simplistic for this dataset.

\begin{table}[ht]
\begin{center}
\caption{Carcinogenic Toxicity of Chemicals Data Analysis (m=4): Parameter estimates, robust standard error estimates, and the robust $p$-value for each main and interaction effects in QEBD models. The QEBD$_F$ is the full model, QEBD$_R$ is the reduced model resulting from backward elimination with QIC, and QELR-CI is the model with the common interaction effect $\gamma$. \label{tb:CTC}}
\begin{tabular}{@{\extracolsep\fill}rrrrrrrrrr@{\extracolsep\fill}}
  \hline
 & \multicolumn{3}{c}{QEBD$_F$} & \multicolumn{3}{c}{QEBD$_R$} & \multicolumn{3}{c}{QELR-CI}\\
 \cmidrule{2-4} \cmidrule{5-7} \cmidrule{8-10}
 & Est. & s.e. & $p$-val. & Est. & s.e. & $p$-val. & Est. & s.e. & $p$-val.\\ 
  \hline
  \rowcolor{lightgray} \multicolumn{10}{l}{Main Effects} \\
  SAL & -3.918 &  1.102 & 0.000 & -3.844 & 1.228 & 0.002 & -4.043  & 0.642 & 0.000\\ 
  MLA & -1.056 &  0.434 & 0.015 & -1.022 & 0.425 & 0.016 & -0.787  & 0.303 & 0.009\\ 
  ABS & -2.619 &  0.699 & 0.000 & -2.450 & 0.594 & 0.000 & -3.177  & 0.564 & 0.000\\ 
  SCE & -1.407 &  0.487 & 0.004 & -1.434 & 0.469 & 0.002 & -0.988  & 0.308 & 0.001\\ 
  \rowcolor{lightgray} \multicolumn{10}{l}{Interaction Effects}\\
  SAL-MLA & 2.465  & 1.348 & 0.068 & 2.595 & 1.131 & 0.022\\ 
  SAL-ABS & 1.865  & 0.589 & 0.002 & 1.930 & 0.538 & 0.000\\ 
  SAL-SCE & 0.275  & 0.872 & 0.752 \\ 
  MLA-ABS & 0.341  & 0.830 & 0.681 \\ 
  MLA-SCE & 2.421  & 0.637 & 0.000 & 2.517 & 0.576 & 0.000\\ 
  ABS-SCE & 2.019  & 0.715 & 0.005 & 2.123 & 0.666 & 0.001\\ 
  C. Interaction $\gamma$ &      &       &    &   &    &   & 1.566 & 0.222 & 0.000 \\ 
   \hline
   \multicolumn{1}{l}{QIC} & \multicolumn{3}{c}{358.349} & \multicolumn{3}{c}{348.900} & \multicolumn{3}{c}{349.940}\\
   \hline
\end{tabular}\end{center}
\end{table}

\subsection{Constitutional Court Opinion Writing Among Justices}

In the realm of appellate court proceedings, the final verdict results from a complex series of decisions made by judges throughout the life of a case. Rather than functioning in isolation, judges participate in a collaborative process with their colleagues to formulate a judicial opinion that encapsulates the collective perspective of the court. This interaction among judges plays a pivotal role in shaping the outcomes they deliver. Judges who agree with the decision but have differing legal interpretations may choose to join or author a concurring opinion. Conversely, those who oppose both the decision and the majority’s legal reasoning have the option to align with or compose a dissenting opinion.

Previous research has explored both individual factors that influence judges’ voting behaviors and the emergence of non-consensual opinions (\citealp{Revesz1997,farhang2004,Peresie2005,boyd2010,Hall2016,Ward2023}), and the impact of peer interactions on judges’ decisions and opinions (\citealp{Wahlbeck1999,Zorn2001,fischman2015,Holden2021}). The applied statistical models included but were not limited to logistic regression, partial proportional odds model, autoregressive (Markov) model, GEE, and nonlinear models. An obvious gap in the aforementioned research is the integration of both the individual and the interaction models. To address this gap, the present study explores the extent to which justices' social networks—rooted in shared educational or professional experiences—influence their propensity to align with one another's opinions.

We hypothesize that justices' social networks, particularly shared educational or professional backgrounds, significantly influence their propensity to align with one another's opinions. This hypothesis builds on the observation that justices begin by assessing the issue at hand and the case outcome based on collective votes and prevailing rationales. They then consider their colleagues' perspectives before deciding to adopt a concurring or dissenting position. These case-specific issues and outcomes represent key factors that justices leverage while collaboratively constructing non-consensual opinions. As their tenure within the court progresses, justices become increasingly familiar with one another, thereby enhancing their collaborative decision-making processes.

In this study, we apply QELR to the Taiwan Constitutional Court dataset as it allows simultaneous modeling of individual justice effects and dyadic interactions within a single framework. While traditional logistic regression treats observations as independent, QELR accounts for the interdependence structure inherent in judicial panel decisions where the same justices appear across multiple cases.

Our analysis focuses on the October 2016–September 2019 term, a period representing a stable composition of the court with all 15 justices serving throughout, eliminating the need to account for membership changes. While this temporal restriction limits our sample to 344 opinions, it ensures that observed interaction patterns reflect genuine justice-to-justice dynamics rather than compositional artifacts. We coded the following variables as main effects: contributing justices (a 15-level categorical variable), issue type (constitutional rights, constitutional institutions, or legal rights), case outcome (constitutional ruling or unconstitutional ruling), and justices' tenure length in years. For interaction effects, we examined justices' educational backgrounds—whether pairs of justices both obtained foreign degrees (from either common-law or civil-law countries) or neither—and prior professional experiences, defined by whether pairs of justices shared the same occupation (academic or legal).

We recognize that our binary categorizations of educational background and professional experience represent substantial simplifications of complex career trajectories. These operational definitions were chosen to maintain adequate cell sizes for analysis given our sample constraints. Specifically, with 105 possible justice pairs and 344 opinions, more granular categorizations would result in sparse cells and unstable estimates. Given data availability constraints common in judicial research, this study adopts an exploratory rather than confirmatory approach. Variable selection was performed using backward elimination with QIC; we acknowledge that stepwise procedures may produce optimistic estimates and limit generalizability. Table~\ref{tb:GJD} presents the QELR results. While issue type, case outcome, and tenure showed no statistically significant effects at conventional levels, the interaction between justices' educational backgrounds revealed an unexpected pattern. The absence of statistical significance for these main effects does not imply the absence of practical significance, particularly for tenure effects which showed a trend toward positive association.

Contrary to our initial hypothesis, shared educational background showed a significant negative association with opinion alignment. This unexpected finding warrants careful interpretation. One possibility is a “distinction-seeking” behavior where justices with similar training deliberately differentiate their jurisprudential positions to establish unique judicial identities. Alternatively, this could indicate that educational diversity within opinion coalitions strengthens legal arguments by incorporating varied jurisprudential traditions. However, given our limited sample and exploratory analysis approach, this finding requires replication before drawing firm theoretical conclusions. Future research with larger samples or longer time periods could explore more nuanced categorizations and test the robustness of these patterns.

\begin{table}
\begin{center}
\caption{Grand Justice Data Analysis ($m=15$): Parameter estimates, robust standard error estimates, and the robust $p$-value for each main and interaction effects in QELR models. The QELR$_F$ is the full model, and QELR$_R$ is the reduced model resulting from backward elimination with QIC.\label{tb:GJD}}
\begin{tabular}{@{\extracolsep\fill}llrrlrrl@{\extracolsep\fill}}
\hline
            & & \multicolumn{3}{c}{QELR$_F$ (QIC=2781.2)} & \multicolumn{3}{c}{QELR$_R$ (QIC=2769.5)}\\
            \cmidrule{3-5} \cmidrule{6-8}
            & & Est. & s.e. & $p$-value & Est. & s.e. & $p$-value\\
            \midrule
\rowcolor{lightgray} \multicolumn{8}{l}{Main Effects}\\ 
Judge ID \\
\multicolumn{2}{r}{GJ1}  &  -2.345 & 0.349 & 0.000 & -2.422 & 0.208 & 0.000\\
\multicolumn{2}{r}{GJ2}  &  -2.577 & 0.355 & 0.000 & -2.644 & 0.247 & 0.000\\
\multicolumn{2}{r}{GJ3}  &  -1.626 & 0.327 & 0.000 & -1.703 & 0.178 & 0.000\\
\multicolumn{2}{r}{GJ4}  &  -1.940 & 0.372 & 0.000 & -1.903 & 0.179 & 0.000\\
\multicolumn{2}{r}{GJ5}  &  -1.773 & 0.355 & 0.000 & -1.739 & 0.168 & 0.000\\
\multicolumn{2}{r}{GJ6}  &  -1.720 & 0.342 & 0.000 & -1.799 & 0.194 & 0.000\\ 
\multicolumn{2}{r}{GJ7}  &  -3.422 & 0.443 & 0.000 & -3.537 & 0.383 & 0.000\\
\multicolumn{2}{r}{GJ8}  &  -3.786 & 0.478 & 0.000 & -3.876 & 0.412 & 0.000\\
\multicolumn{2}{r}{GJ9}  &  -1.627 & 0.286 & 0.000 & -1.730 & 0.181 & 0.000\\
\multicolumn{2}{r}{GJ10} &  -1.573 & 0.306 & 0.000 & -1.675 & 0.184 & 0.000\\
\multicolumn{2}{r}{GJ11} &  -2.383 & 0.397 & 0.000 & -2.345 & 0.200 & 0.000\\
\multicolumn{2}{r}{GJ12} &  -2.332 & 0.318 & 0.000 & -2.433 & 0.237 & 0.000\\
\multicolumn{2}{r}{GJ13} &  -1.663 & 0.310 & 0.000 & -1.767 & 0.190 & 0.000\\
\multicolumn{2}{r}{GJ14} &  -2.601 & 0.371 & 0.000 & -2.555 & 0.219 & 0.000\\
\multicolumn{2}{r}{GJ15} &  -1.537 & 0.320 & 0.000 & -1.617 & 0.162 & 0.000\\
Issue\\
\multicolumn{2}{r}{Const. Rights} & -0.357 & 0.233 &  0.125 \\
\multicolumn{2}{r}{\hspace{.3in} Const. Institutions} & -0.408 & 0.247 & 0.099\\
Case Outcome\\
\multicolumn{2}{r}{Const. Ruling} & 0.270 & 0.197 &  0.170\\
\multicolumn{2}{r}{Unconst. Ruling} & 0.293 & 0.204 & 0.150 \\
Time ($t$)\\
\multicolumn{2}{r}{$t$}      &   -1.078 & 2.354 & 0.647 \\
\multicolumn{2}{r}{$t^2$}    &   3.741  & 6.203 & 0.546 \\
\multicolumn{2}{r}{$t^3$}    &   -2.753 & 3.878 & 0.478 \\
& \\
\rowcolor{lightgray}  \multicolumn{8}{l}{Interaction Effects}\\ 
\multicolumn{2}{r}{Prior Occupation} & 0.065 & 0.254 & 0.797\\
\multicolumn{2}{r}{Education} & -1.288 & 0.247 & 0.000 &-1.240 & 0.253 & 0.000 \\
\hline
\end{tabular}\end{center}
\end{table}

\section{Conclusion}

This work explores the estimation challenges of conditional mean models for correlated binary response variables within longitudinal data and network data contexts. Traditional methods such as GPL and GEE are scrutinized for their potential pitfalls when applied without sufficient caution. In particular, we prove that GEE with independence working correlation guarantees estimation consistency, but GEE with other widely used alternatives, such as compound symmetry and autoregressive correlations, do not. We hope to draw the attention of the researchers to carefully consider their methodological choices in conditional mean models to ensure the accuracy and reliability of statistical analyses. Moreover, although we focus on multivariate binary response variables, Theorem~\ref{th:gee} holds for all variables belonging to the Exponential family. 

The conditional mean models resulting from the QEBD and QELR have exactly the form of logistic regressions; hence, the model inherits the pros and cons of the logistic regression. \cite{Firth1993} pointed out that some true parameter values do not exist when the data is separable. Imposing certain penalty terms, such as $l_1$ and/or $l_2$,  is overwhelmingly welcomed when $m$ is mild to large, \citet{Canditiis2020}. Additionally, using GPL raises another computing issue. Consider a dataset comprising $n$ samples, each associated with $m$ binary responses, for example. The resulting design matrix of the GPL comprises $n \times m$ rows. This size of the design matrix can potentially lead to computer memory overflow. Fortunately, this problem has been properly resolved in terms of solving GLMs. \cite{enea2009} has proposed strategies for handling large datasets, and \cite{Wang2025} has proposed online algorithms for high-dimensional data. Adapting and integrating these strategies into GEE computations is essential for solving large-$m$ GPL by GEE. We defer the implementation to our future study.

\backmatter

\section*{Acknowledgements}



\bibliographystyle{biom} \bibliography{GJ}

\begin{thebibliography}{}

\bibitem[\protect\citeauthoryear{Agresti}{Agresti}{2019}]{Agresti2019}
Agresti, A. (2019).
\newblock {\em An Introduction to Categorical Data Analysis}.
\newblock John Wiley \& Sons Inc, New Jersey, third edition.

\bibitem[\protect\citeauthoryear{Akaike}{Akaike}{1973}]{Akaike1973}
Akaike, H. (1973).
\newblock Information theory and an extension of the maximum likelihood principle.
\newblock {\em In Proceedings of the Second International Symposium on Information Theory, B. N. Petrov and F. Csaki (eds)} pages 267--281.

\bibitem[\protect\citeauthoryear{Arnold and Press}{Arnold and Press}{1989}]{Arnold1989}
Arnold, B. and Press, S. (1989).
\newblock Compatible conditional distributions.
\newblock {\em Journal of the American Statistical Association} {\bf 84,} 152--156.

\bibitem[\protect\citeauthoryear{Bible, Albert, Simons-Morton, and Liu}{Bible et~al.}{2019}]{Bible2019}
Bible, J., Albert, P.~S., Simons-Morton, B.~G., and Liu, D. (2019).
\newblock Practical issues in using generalized estimating equations for inference on transitions in longitudinal data: What is being estimated?
\newblock {\em Statistics in Medicine} {\bf 38,} 903--916.

\bibitem[\protect\citeauthoryear{Bishop, Fienberg, and Holland}{Bishop et~al.}{1975}]{Bishop1975}
Bishop, Y. M.~M., Fienberg, S.~E., and Holland, P.~W. (1975).
\newblock {\em Discrete multivariate analysis: theory and practice}.
\newblock MIT Press, Cambridge, Massachussetts.

\bibitem[\protect\citeauthoryear{Boyd, Epstein, and Martin}{Boyd et~al.}{2010}]{boyd2010}
Boyd, C.~L., Epstein, L., and Martin, A.~D. (2010).
\newblock Untangling the causal effects of sex on judging.
\newblock {\em American Journal of Political Science} {\bf 54,} 389--411.

\bibitem[\protect\citeauthoryear{Brusco, Steinley, and Watts}{Brusco et~al.}{2023}]{Brusco2023}
Brusco, M.~J., Steinley, D., and Watts, A. (2023).
\newblock A comparison of logistic regression methods for ising model estimation.
\newblock {\em Behavior Research Methods} {\bf 53,} 3566--3584.

\bibitem[\protect\citeauthoryear{Connolly and Liang}{Connolly and Liang}{1988}]{Connolly1988}
Connolly, M. and Liang, K. (1988).
\newblock Conditional logistic regression models for correlated binary data.
\newblock {\em Biometrika} {\bf 75,} 501--506.

\bibitem[\protect\citeauthoryear{Cox}{Cox}{1972}]{Cox1972}
Cox, D. (1972).
\newblock The analysis of multivariate binary data.
\newblock {\em Journal of the Royal Statistical Society. Series C} {\bf 21,} 113--120.

\bibitem[\protect\citeauthoryear{Cox and Wermuth}{Cox and Wermuth}{1994}]{Cox1994}
Cox, D.~R. and Wermuth, N. (1994).
\newblock A note on the quadratic exponential binary distribution.
\newblock {\em Biometrika} {\bf 81,} 403--408.

\bibitem[\protect\citeauthoryear{De~Canditiis}{De~Canditiis}{2020}]{Canditiis2020}
De~Canditiis, D. (2020).
\newblock A global approach for learning sparse ising models.
\newblock {\em Mathematics and Computers in Simulation} {\bf 176,} 160–--170.

\bibitem[\protect\citeauthoryear{Diggle, Heagerty, Liang, and Zeger}{Diggle et~al.}{2002}]{Diggle2002}
Diggle, P.~J., Heagerty, P., Liang, K.-Y., and Zeger, S.~L. (2002).
\newblock {\em Analysis of Longitudinal Data}.
\newblock Oxford University Press, Oxford, second edition.

\bibitem[\protect\citeauthoryear{Enea}{Enea}{2009}]{enea2009}
Enea, M. (2009).
\newblock Fitting linear models and generalized linear models with large data sets in r.
\newblock {\em Statistical Methods for the Analysis of Large Datasets: book of short papers} pages 411--414.

\bibitem[\protect\citeauthoryear{Farhang and Wawro}{Farhang and Wawro}{2004}]{farhang2004}
Farhang, S. and Wawro, G. (2004).
\newblock Institutional dynamics on the u.s. court of appeals: Minority representation under panel decision making.
\newblock {\em The Journal of Law, Economics, and Organization} {\bf 20,} 299--330.

\bibitem[\protect\citeauthoryear{Firth}{Firth}{1993}]{Firth1993}
Firth, D. (1993).
\newblock Bias reduction of maximum likelihood estimates.
\newblock {\em Biometrika} {\bf 80,} 27--38.

\bibitem[\protect\citeauthoryear{Fischman}{Fischman}{2015}]{fischman2015}
Fischman, J.~B. (2015).
\newblock Interpreting circuit court voting patterns: A social interactions framework.
\newblock {\em The Journal of Law, Economics, and Organization} {\bf 31,} 808--842.

\bibitem[\protect\citeauthoryear{Halekoh, Højsgaard, and Yan}{Halekoh et~al.}{2006}]{geepack2006}
Halekoh, U., Højsgaard, S., and Yan, J. (2006).
\newblock The r package geepack for generalized estimating equations.
\newblock {\em Journal of Statistical Software} {\bf 15/2,} 1--11.

\bibitem[\protect\citeauthoryear{Hall and Windett}{Hall and Windett}{2016}]{Hall2016}
Hall, M. E.~K. and Windett, J.~H. (2016).
\newblock Discouraging dissent: The chief judge’s influence in state supreme courts.
\newblock {\em American Politics Research} {\bf 44,} 682--709.

\bibitem[\protect\citeauthoryear{Haseman, Zeiger, Shelby, Margolin, and Tennant}{Haseman et~al.}{1990}]{Haseman1990}
Haseman, J.~K., Zeiger, E., Shelby, M.~D., Margolin, B.~H., and Tennant, R.~W. (1990).
\newblock Predicting rodent carcinogenicity from four in vitro genetic toxicity assays: An evaluation of 114 chemicals studied by the national toxicology program.
\newblock {\em Journal of the American Statistical Association} {\bf 85,} 964--971.

\bibitem[\protect\citeauthoryear{Hastie, Tibshirani, and Friedman}{Hastie et~al.}{2009}]{Hastie2009}
Hastie, T., Tibshirani, R., and Friedman, J. (2009).
\newblock {\em The Elements of Statistical Learning: Data Mining, Inference, and Prediction}.
\newblock Springer, New York, second edition.

\bibitem[\protect\citeauthoryear{Holden, Keane, and Lilley}{Holden et~al.}{2021}]{Holden2021}
Holden, R., Keane, M., and Lilley, M. (2021).
\newblock Peer effects on the united states supreme court.
\newblock {\em Quantitative Economics} {\bf 12,} 981--1019.

\bibitem[\protect\citeauthoryear{Jirousek and Preucil}{Jirousek and Preucil}{1995}]{Jirousek1995}
Jirousek, R. and Preucil, S. (1995).
\newblock On the effective implementation of the iterative proportional fitting procedure.
\newblock {\em Computational Statistics and Data Analysis} {\bf 19,} 177--189.

\bibitem[\protect\citeauthoryear{Joe and Liu}{Joe and Liu}{1996}]{Joe1996}
Joe, H. and Liu, Y. (1996).
\newblock A model for a multivariate binary response with covariates based on compatible conditionally specified logistic regression.
\newblock {\em Statistics \& Probability Letters} {\bf 31,} 113--120.

\bibitem[\protect\citeauthoryear{Liang and Zeger}{Liang and Zeger}{1986}]{Liang1986}
Liang, K. and Zeger, S. (1986).
\newblock Longitudinal data analysis using generalized linear models.
\newblock {\em Biometrika} {\bf 73,} 13--22.

\bibitem[\protect\citeauthoryear{Lipsitz and Fitzmaurice}{Lipsitz and Fitzmaurice}{1994}]{Lipsitz1994}
Lipsitz, S.~R. and Fitzmaurice, G. (1994).
\newblock An extension of yule's q to multivariate binary data.
\newblock {\em Biometrics} {\bf 50,} 847--852.

\bibitem[\protect\citeauthoryear{McCullagh and Nelder}{McCullagh and Nelder}{1983}]{McCullagh1983}
McCullagh, P. and Nelder, J.~A. (1983).
\newblock {\em Generalized Linear Models}.
\newblock Chapman and Hall.

\bibitem[\protect\citeauthoryear{Myers, Montgomery, Vining, and Robinson}{Myers et~al.}{2010}]{Myers2010}
Myers, R.~H., Montgomery, D.~C., Vining, G.~G., and Robinson, T.~J. (2010).
\newblock {\em Generalized Linear Models with Applications in Engineering and the Sciences}.
\newblock A John Wiley \& Sons, INC., 2 edition.

\bibitem[\protect\citeauthoryear{Pan}{Pan}{2001}]{Pan2001}
Pan, W. (2001).
\newblock Akaike's information criterion in generalized estimating equations.
\newblock {\em Biometrics} {\bf 57,} 120--125.

\bibitem[\protect\citeauthoryear{Pan and Connett}{Pan and Connett}{2002}]{Pan2002}
Pan, W. and Connett, J. (2002).
\newblock Selecting the working correlation structure in generalized estimating equations with application to the lung health study.
\newblock {\em Statistica Sinica} {\bf 12,} 475--490.

\bibitem[\protect\citeauthoryear{Pan, Louis, and Connett}{Pan et~al.}{2000}]{Pan2000}
Pan, W., Louis, T.~A., and Connett, J.~E. (2000).
\newblock A note on marginal linear regression with correlated response data.
\newblock {\em The American Statistician} {\bf 54,} 191--195.

\bibitem[\protect\citeauthoryear{Pepe and Anderson}{Pepe and Anderson}{1994}]{Pepe1994}
Pepe, M. and Anderson, G. (1994).
\newblock A cautionary note on inference for marginal regression models with longitudinal data and general correlated response data.
\newblock {\em Communications in Statistics - Simulation and Computation} {\bf 23,} 939--951.

\bibitem[\protect\citeauthoryear{Peresie}{Peresie}{2005}]{Peresie2005}
Peresie, J.~L. (2005).
\newblock Female judges matter: Gender and collegial decisionmaking in the federal appellate courts.
\newblock {\em The Yale Law Journal} {\bf 114,} 1759--1790.

\bibitem[\protect\citeauthoryear{Peterson and Anderson}{Peterson and Anderson}{1987}]{Peterson1987}
Peterson, C. and Anderson, J.~R. (1987).
\newblock A mean field theory learning algorithm for neural networks.
\newblock {\em Complex Systems} {\bf 1,} 995--1019.

\bibitem[\protect\citeauthoryear{Qu, Williams, Beck, and Goormastic}{Qu et~al.}{1987}]{Qu1987}
Qu, Y., Williams, G., Beck, G., and Goormastic, M. (1987).
\newblock A generalized model of logistic regression for correlated data.
\newblock {\em Communication of Statistics, A.} {\bf 16,} 3447--3476.

\bibitem[\protect\citeauthoryear{{R Core Team}}{{R Core Team}}{2021}]{R2021}
{R Core Team} (2021).
\newblock {\em R: A Language and Environment for Statistical Computing}.
\newblock R Foundation for Statistical Computing, Vienna, Austria.

\bibitem[\protect\citeauthoryear{Ravikumar, Wainwright, and Lafferty}{Ravikumar et~al.}{2010}]{Ravikumar2010}
Ravikumar, P., Wainwright, M., and Lafferty, J. (2010).
\newblock High-dimensional {I}sing model selection using $l_1$-regularized logistic regression.
\newblock {\em The Annals of Statistics} {\bf 38,} 1287--1319.

\bibitem[\protect\citeauthoryear{Revesz}{Revesz}{1997}]{Revesz1997}
Revesz, R.~L. (1997).
\newblock Environmental regulation, ideology, and the d. c. circuit.
\newblock {\em Virginia Law Review} {\bf 83,} 1717--1772.

\bibitem[\protect\citeauthoryear{Ripley}{Ripley}{1996}]{Ripley1996}
Ripley, B.~D. (1996).
\newblock {\em Pattern Recognition and Neural Networks}.
\newblock Cambridge University Press.

\bibitem[\protect\citeauthoryear{Stefanski and Boos}{Stefanski and Boos}{2002}]{Stefanski2002}
Stefanski, L. and Boos, D. (2002).
\newblock The calculus of m-estimation.
\newblock {\em The American Statistician} {\bf 56,} 29--38.

\bibitem[\protect\citeauthoryear{Strauss and Ikeda}{Strauss and Ikeda}{1990}]{Strauss1990}
Strauss, D. and Ikeda, M. (1990).
\newblock Pseudolikelihood estimation for social networks.
\newblock {\em Journal of the American Statistical Association} {\bf 85,} 204--212.

\bibitem[\protect\citeauthoryear{Wahlbeck, James F.~Spriggs, and Maltzman}{Wahlbeck et~al.}{1999}]{Wahlbeck1999}
Wahlbeck, P.~J., James F.~Spriggs, I., and Maltzman, F. (1999).
\newblock The politics of dissents and concurrences on the u.s. supreme court.
\newblock {\em American Politics Quarterly} {\bf 27,} 488--514.

\bibitem[\protect\citeauthoryear{Wang, Wei, and Yao}{Wang et~al.}{2025}]{Wang2025}
Wang, X., Wei, M.~M., and Yao, T. (2025).
\newblock Online learning and decision making under generalized linear model with high-dimensional data.
\newblock {\em MANAGEMENT SCIENCE} {\bf 71,} 6647--6665.

\bibitem[\protect\citeauthoryear{Ward, Corley, and Steigerwalt}{Ward et~al.}{2023}]{Ward2023}
Ward, A., Corley, P.~C., and Steigerwalt, A. (2023).
\newblock {\em The Puzzle of Unanimity: Consensus on the United States Supreme Court}.
\newblock Stanford University Press, Stanford.

\bibitem[\protect\citeauthoryear{Zeger and Qaqish}{Zeger and Qaqish}{1988}]{Zeger1988}
Zeger, S.~L. and Qaqish, B. (1988).
\newblock Markov regression models for time series: A quasi-likelihood approach.
\newblock {\em Biometrics} {\bf 44,} 1019--1031.

\bibitem[\protect\citeauthoryear{Zhao and Prentice}{Zhao and Prentice}{1990}]{Zhao1990}
Zhao, L. and Prentice, R. (1990).
\newblock Correlated binary regression using a quadratic exponential model.
\newblock {\em Biometrika} {\bf 77,} 642--648.

\bibitem[\protect\citeauthoryear{Zorn}{Zorn}{2001}]{Zorn2001}
Zorn, C. J.~W. (2001).
\newblock Generalized estimating equation models for correlated data: A review with applications.
\newblock {\em American Journal of Political Science} {\bf 45,} 470--490.

\end{thebibliography}

\section*{Supporting Information}

\appendix

\section{}

\subsection{Proof of Theorem~\ref{th:gee}} 
\begin{proof}
    Denote $C_j = E\left\{ \mathbf Y_{[j]}^{0}(\mathbf Y-\boldsymbol \mu)^{\top}\nu_j\right\} \in \mathbb{R}^{m\times m}$ where $\mu_j=E(Y_j|\mathbf Y_{[j]})$ and $\nu_j=Var(Y_j|\mathbf Y_{[j]})$. So both $\mu_j$ and $\nu_j$ are independent of $Y_j$. Moreover, for $k \neq j$,
    \[
        [C_j]_{kj} = E\{ Y_k (Y_j-\mu_j)\nu_j\} = E_{Y_{[j]}}\left\{Y_k \nu_j E_{Y_j|Y_{[j]}}(Y_j-\mu_j) \right\}=0
    \]
    because of the double expectation rule. Furthermore, $[C_j]_{jk} = 0\times E\{(Y_k-\mu_k)\nu_j\}$, $k=1,\dots,m$. So $C_j$ is a square matrix with its $j$th column and $j$th row equal to an $m$-dimensional zero vector.

    Equation (\ref{eq:gee_main}) can be expressed as  
    \[
        \boldsymbol \varphi(\boldsymbol \theta) =\sum_{j=1}^m \left[ \begin{array}{c} \mathbf x_j \\ W_j \mathbf Y_{[j]}^{0}\end{array}\right] \mathbf e_j^{\top}A V^{-1}(\mathbf Y-\boldsymbol\mu).
    \]
    Since $E(Y_j - \pi_j)=0$, the upper part, corresponding to $\beta$, has mean zero. Moreover, the expectation of the lower part, corresponding to $\gamma$, is
    \[
        E\left[\sum_{j=1}^m W_j \mathbf{Y}_{[j]}^0(\mathbf{Y} - \boldsymbol{\pi})^{\top}V^{-1}\mathbf{e}_j\nu_j\right]
        = \sum_{j=1}^m W_j E\left[ \mathbf{Y}_{[j]}^0(\mathbf{Y} - \boldsymbol{\pi})^{\top}\nu_j\right]V^{-1}\mathbf{e}_j
        = \sum_{j=1}^m W_jC_jV^{-1}\mathbf{e}_j. 
    \]
    Next, when $V$ is a diagonal matrix, $V^{-1}\mathbf e_j = \mathbf e_j [V^{-1}]_{jj}$. Moreover, since $C_j$ is a square matrix whose $j$th column is a zero vector, we conclude that $C_j \mathbf e_j=\mathbf 0$, and hence, $C_jV^{-1}\mathbf e_j=\mathbf 0$. The proof is complete.
\end{proof}

\label{lastpage}

\end{document}